\documentclass[a4paper,11pt]{article}%
\usepackage{amsmath}
\usepackage{graphicx}%
\usepackage{amsfonts}%
\usepackage{amssymb}
\newtheorem{theorem}{Theorem}

\newtheorem{definition}[theorem]{Definition}

\begin{document}

\title{A Unified Scheme for Generalized Sectors based on Selection Criteria. \\I. Thermal situations, unbroken symmetries and criteria as classifying
categorical adjunctions}
\author{Izumi Ojima\\Research Institute for Mathematical Sciences, Kyoto University\\Kyoto 606-8502 Japan}
\date{}
\maketitle

\begin{abstract}
A unified scheme for treating generalized superselection sectors is proposed
on the basis of the notion of selection criteria to characterize states of
relevance to each specific domain in quantum physics, ranging from the
relativistic quantum fields in the vacuum situations with unbroken and
spontaneously broken internal symmetries, through equilibrium and
non-equilibrium states to some basic aspects in measurement processes.
This is achieved by the help of \textit{c} $\rightarrow$ \textit{q} and
\textit{q} $\rightarrow$ \textit{c} channels, the former of which determines
the states to be selected and to be parametrized by the order parameters
defined as the spectrum of the centre constituting the superselection sectors,
and the latter of which provides, as classifying maps, the physical
interpretations of selected states in terms of order parameters. This
formulation extends the traditional range of applicability of the
Doplicher-Roberts construction method for recovering the field algebra and the
gauge group (of the first kind) from the data of group invariant observables
to the situations with spontaneous symmetry breakdown, as will be reported in
a succeeding paper.

\end{abstract}

\section{Introduction}

The stanrdard way of treating the microscopic world on the basis of quantum
field theory (QFT, for short) is to introduce first the \textit{quantum
fields} whose characterization is given by means of their behaviours under the
various kinds of \textit{symmetries}; e.g., the internal symmetry groups such
as the colour $SU(3)$, the chiral $SU(2)$, the electromagnetic $U(1)$, or any
other bigger (super)groups of grand unifications (and their corresponding
versions of local gauge symmetries), in combination with the spacetime
symmetry groups such as the Poincar\'{e} group in the Minkowski spacetime,
conformal groups in massless theories, or isometry groups of curved
spacetimes, and so on. In a word, the basic objects of such a system can
essentially be found in an algebra $\mathfrak{F}$ of quantum fields (called a
\textit{field algebra}, for short) acted upon by two kinds of symmetries,
\textit{internal }and \textit{spacetime} (whose unification has been pursued
as one of the ultimate goals of microscopic physics). With respect to the
group of an internal symmetry denoted generically by $G$, the generating
elements of $\mathfrak{F}$ (usually called \textit{basic} or
\textit{fundamental fields}) are assumed (\textit{by hand}) to belong to
certain multiplet(s) transforming covariantly under the action of $G$, which
defines mathematically an action $\tau$ of $G$ on $\mathfrak{F}$:
$G\underset{\tau}{\curvearrowright}\mathfrak{F}$.

Contrary to this kind of \textit{theoretical }setting, what is to be observed
(experimentally) in the real world is believed (or, can be proved in a certain
setup; see, for instance, \cite{Oji78}) to be only those elements of
$\mathfrak{F}$ \textit{invariant under }$G$ and are usually called
\textit{observables} which constitute the algebra $\mathfrak{A}$ of
observables:
\begin{equation}
\mathfrak{A}:=\mathfrak{F}^{G},
\end{equation}
the fixed-point subalgebra of $\mathfrak{F}$ under the action $\tau$ of $G$.
Thus, what we can directly check experimentally is supposed to be only those
data described in terms of $\mathfrak{A}$ (and its derived objects) and the
rest of the notions appearing in our framework are just mathematical devices
whose pertinence can be justified only through the information related to
$\mathfrak{A}$. Except for the systematic approaches \cite{DHR, DR90, Haag}
undertaken by the group of algebraic QFT, however, there have so far been no
serious attempts to understand the basic mechanism pertaining to this point as
to how a particular choice of $\mathfrak{F}$ and $G$ can be verified, leaving
aside the problems of this sort just to the heuristic arguments based on
trials and errors. While a particularly chosen combination $G\underset{\tau
}{\curvearrowright}\mathfrak{F}$ becomes no doubt meaningless without good
agreements of its consequences with the observed data described in terms of
$\mathfrak{A}$, the attained agreements support the postulated theoretical
assumption only as \textit{one of many possible candidates} of explanations,
without justifying it as a \textit{unique} inevitable solution. (Does it not
look quite strange that such a kind of problems as this have hardly been
examined in the very sophisticated discussions about the unicity of the
unification models at the Planck scale?)

Just when restricted to the cases with $G$ an \textit{unbroken} global gauge
symmetry (or, gauge symmetry of the first kind), a satisfactory framework in
this context has been established in \cite{DR89, DR90}, whose physical essence
has, unfortunately, not been recognized widely (which may be partly due to its
mathematical sophistications, but mainly due to the lack of common
understanding of the importance of the above-mentioned problem). This theory
enables one to \textit{recover both} $\mathfrak{F}$ \textit{and} $G$ starting
only from the data encoded in $\mathfrak{A}$ when supplemented by the
so-called DHR \textit{selection criterion} \cite{Borch, DHR} to pick up
physically relevant states with localizable charges (which should, in the case
of topological charges, be replaced by the one proposed in
Buchholz-Fredenhagen \cite{BF82}). Then, the vacuum representation of the
constructed field algebra $\mathfrak{F}$ is decomposed into mutually disjoint
irreducible representations of $\mathfrak{A}=\mathfrak{F}^{G}$, called
\textit{superselection sectors},\textit{\ }in one-to-one correspondence with
mutually disjoint irreducible unitary representations of the internal symmetry
group $G$ which is found to be \textit{compact Lie}. However, the traditional
notion of sector structure, hinging strongly to the essential features of
unbroken symmetry, has so far allowed only the \textit{discrete sectors},
parametrized by the discrete $\hat{G}$, the dual of a compact group defined as
the set of all equivalence classes of finite-dimensional continuous unitary
irreducible representations of $G$. When we start to extend this formalism to
the situations with \textit{spontaneous symmetry breakdown} (SSB, for short),
we encounter the presence of \textit{continuous sectors} (or, ``degenerate
vacua'' in the traditional terminology) parametrized by continuous
\textit{macroscopic order parameters}, as is seen in the second paper of this
series (\cite{II}, called II hereafter).

Aside from such very fundamental issues as the ``ultimate'' unifications, we
have so far faced with so many different levels and areas in the directions
from microscopic worlds to macroscopic ones, ranging from the vacuum
situations (the standard QFT relevant to particle physics), thermal equilibria
(QFT at finite temperatures or quantum statistical mechanics), non-equilibrium
ones and so on. In \cite{BOR01}, a general framework is proposed for defining
non-equilibrium local states in relativistic QFT and for describing their
thermodynamic properties in terms of the associated macroscopic observables
found in the \textit{centre} of relevant representations of observables. From
the general standpoint, one easily notices that the thermal equilibria at
different temperatures can also be seen to constitute families of continuous
sectors parametrized by such thermodynamic variables as temperatures, chemical
potentials and pressure and so on. In view of such roles of central
observables associated with continuous sectors appearing in SSB cases as well
as the above various kinds of thermal states, it seems appropriate to extend
the notion of sectors so as to incorporate and to try the possibility of
unified ways of treating these different cases, just regarding the traditional
discrete ones as special cases; this is simply parallel to the extension of
the traditional \textit{eigenvalue} problems for linear operators with
\textit{discrete spectra }to the general \textit{spectral decompositions
}admitting the appearance of \textit{continuous spectra}.

The aim of the present series of articles (in combination with \cite{II}) is
to propose a scheme to unify such a generalized notion of sectors from the
viewpoint of the key roles played by the \textit{selection criteria }at the
starting point of theory in defining and choosing \textit{physically relevant
family of states} as well as in providing a systematic way for
\textit{describing} and \textit{interpreting }the relevant physical
properties. In this paper (called I), we introduce the necessary ingredients
for formulating the scheme through the discussions on the basic structures
found in thermal situations of equilibrium (Sec.2) and of the extension to
non-equilibrium (Sec.3) and in the reformulated version of DHR superselection
theory (Sec.4). At the end, we explain the general mathematical meaning of the
proposed scheme, in relation with the categorical adjunctions, especially with
the geometric notions of classifying spaces and classifying maps. The analysis
is to be continued in the second paper \cite{II}, where we develop a
systematic treatment of spontaneously broken symmetries from the viewpoint of
proposed scheme and close the series by examining the operational meaning of
our basic ingredients in relation with certain basic notions relevant to the
quantum measurements.

\section{Equilibrium states and thermal interpretations: roles of
\textit{c}$\rightarrow$\textit{q} and \textit{q}$\rightarrow$\textit{c} channels}

To draw a clear picture of the idea, we briefly sketch the essense of the
scheme proposed in \cite{BOR01} for defining and describing non-equilibrium
local states in a relativistic QFT. From the present standpoint, it can be
reformulated as follows according to \cite{Oji02}. To characterize an unknown
state $\omega$ as a non-equilibrium local state, we prepare the following
basic ingredients.

\begin{itemize}
\item[i)] The set $\mathcal{K}$ of \textit{thermal reference states
}consisting of all global thermal equilibrium states defined as the
relativistic KMS states $\omega_{\beta}$ \cite{BrBu} (with inverse temperature
4-vectors $\beta=(\beta^{\mu})\in V_{+}:=\{x\in\mathbb{R}^{4};x^{0}%
>0,x^{2}=(x^{0})^{2}-\vec{x}^{2}>0\}$) and their suitable convex combinations:
$\mathcal{K}$ plays the role of a \textit{model space }whose analogue in the
definition of a manifold $M$ can be\textit{\ }found in a Euclidean space
$\mathbb{R}^{n}$ as the value space of local charts.

\item[ii)] The set $\mathcal{T}_{x}$ of \textit{local thermal observables}
consisting of suitable combinations of quantum observables at a spacetime
point $x$ defined mathematically as linear forms on states with suitable
regularity in their energy contents (which are quantum observables in an
extended sense) \cite{Bo, BOR01}. Along the above analogy to a manifold $M$ in
differential geometry, their role is to relate our unknown state $\omega$ to
the known reference states in $\mathcal{K}$, just in parallel to the local
coordinates which relate locally a generic curved space $M$ to the known space
$\mathbb{R}^{n}$. As explained just below, the physical interpretations of
local thermal observables $\hat{A}$ are given by macroscopic \textit{thermal
functions} corresponding to $\hat{A}$, through which our unknown $\omega$ can
be \textit{compared} with thermal reference states in $\mathcal{K}$.
\end{itemize}

Before going into the discussion of non-equilibrium, we need first to
establish the physical roles of the above ingredients for describing the
relevant thermal properties of states and quantum observables in the realm
$\mathcal{K}$ of generalized thermal equilibria. To this end, we introduce

\begin{definition}
\textit{\textbf{Thermal functions }}are defined for each quantum observables
$\hat{A}$($\in\mathcal{T}_{x}$) by the map
\begin{align}
\Xi & :\hat{A}\longmapsto\Xi(\hat{A})\in C(B_{\mathcal{K}})\text{
\ }\nonumber\\
& \text{with\ }B_{\mathcal{K}} \ni(\beta,\mu)\longmapsto\Xi(\hat{A})(\beta
,\mu):=\omega_{\beta,\mu}(\hat{A}),
\end{align}
where $B_{\mathcal{K}}$ is the classifying space to parameterize thermodynamic
pure phases, consisting of inverse temperature 4-vectors $\beta\in V_{+}$ in
addition to any other thermodynamic parameters (if any) generically denoted by
$\mu$ (e.g., chemical potentials) necessary to exhaust and discriminate all
the thermodynamic pure phases.
\end{definition}

Since this map $\Xi$ is easily seen to be a unital (completely) positive
linear map, $\Xi(\mathbf{1})=1,\Xi(\hat{A}^{\ast}\hat{A})\geq0$,\thinspace
\ its dual map $\Xi^{\ast}$ on states becomes a \textit{\textbf{c}%
lassical-\textbf{q}uantum (c}$\rightarrow$\textit{q) channel} $\Xi^{\ast
}:M_{1}(B_{\mathcal{K}})\ni\rho\longmapsto\Xi^{\ast}(\rho)\in\mathcal{K}$
given by%
\begin{align}
& \Xi^{\ast}(\rho)(\hat{A})=\rho(\Xi(\hat{A}))=\int_{B_{\mathcal{K}}}%
d\rho(\beta,\mu)\Xi(\hat{A})(\beta,\mu)=\int_{B_{\mathcal{K}}}d\rho(\beta
,\mu)\omega_{\beta,\mu}(\hat{A}),\nonumber\\
& \Longrightarrow\Xi^{\ast}(\rho):=\int_{B_{\mathcal{K}}}d\rho(\beta
,\mu)\omega_{\beta,\mu}=\omega_{\rho}\in\mathcal{K}.
\end{align}
Here $M_{1}(B_{\mathcal{K}})$ is the space of probability measures $\rho$ on
$B_{\mathcal{K}}$ describing the mean values of thermodynamic parameters
$(\beta,\mu)$ together with their fluctuations. One can see that thermal
interpretation of local quantum thermal observables $\hat{A}\in\mathcal{T}%
_{x}$ is given in all thermal reference states of the form $\Xi^{\ast}%
(\rho)=\omega_{\rho}\in\mathcal{K}$ by the corresponding thermal observable
$\Xi(\hat{A})$ evaluated in the classical probability $\rho$ describing the
thermodynamic configurations of $\omega_{\rho}$ through the relation
\begin{equation}
\omega_{\rho}(\hat{A})=\int_{B_{\mathcal{K}}}\!d\rho(\beta,\mu)\,\omega
_{\beta,\mu}(\hat{A})=\int_{B_{\mathcal{K}}}\!d\rho(\beta,\mu)[\,\Xi
(\hat{A})](\beta,\mu)=\rho(\Xi(\hat{A})).
\end{equation}
This applies to the case where $\rho$ is already known. What we need in the
actual situations is how to determine the unknown $\rho$ from the given data
set $\Phi\longmapsto\rho(\Phi)$ of expectation values of thermal functions
$\Phi$ (which is the problem of state estimation): this problem can be solved
if $\mathcal{T}_{x}$ has sufficiently many local thermal observables so that
the totality $\Xi(\mathcal{T}_{x})$ of the corresponding thermal functions can
approximate arbitrary continuous functions of $(\beta,\mu)\in B_{\mathcal{K}}%
$. In this case $\rho$ is given as the unique solution to a (generalized)
``moment problem''. Thus we see:

\begin{itemize}
\item[$\bigstar$] If the set $\mathcal{T}_{x}$\textit{\ }of local thermal
observables are sufficient for discriminating all the thermal reference states
in $\mathcal{K}$, then any reference state $\in\mathcal{K}$ can be written as
$\Xi^{\ast}(\rho)$ in terms of a uniquely determined probability measure
$\rho\in B_{\mathcal{K}}$. Then local thermal observables\textit{\ }$\hat
{\Phi}\in\mathcal{T}_{x}$\textit{\ }provide the same information on the
thermal properties of states in $\mathcal{K}$ as that provided by the
corresponding classical macroscopic thermal functions\textit{\ }$\Phi=\Xi
(\hat{\Phi})$\ [e.g., internal energy, entropy density, etc.]: $\omega_{\rho
}(\hat{\Phi})=\rho(\Phi)$.
\end{itemize}

In this situation, the \textit{inverse} of $\Xi^{\ast}$ becomes meaningful on
$\mathcal{K}$ and the thermal interpretation of thermal reference states
$\in\mathcal{K}$ is just given by this \textbf{\textit{\textbf{q} }%
}$\rightarrow$\textit{\textbf{c channel }}$(\Xi^{\ast})^{-1}$\textbf{\ }%
$:\mathcal{K}\ni\omega\longmapsto\rho\in M_{1}(B_{\mathcal{K}})$ s.t.
$\omega=\Xi^{\ast}(\rho)$ \cite{Oji02}, which can be regarded as a simple
adaptation and extension of the notions of classifying spaces and classifying
maps to the context involving (quantum) probability theory. We express the
essence of the above situation ($\bigstar$) by
\begin{equation}
\mathcal{K}(\omega,\Xi^{\ast}(\rho))/\mathcal{T}_{x}\overset{q\rightleftarrows
c}{\simeq}Th((\Xi^{\ast})^{-1}(\omega),\rho)/\Xi(\mathcal{T}_{x}%
).\label{adjunction1}%
\end{equation}
What is to be noted here is that two levels, quatum statistical mechanics with
family $\mathcal{K}$ of KMS states and macroscopic thermodynamics with
parameter space $B_{\mathcal{K}}$, are interrelated by means of the
channels,\textit{\textbf{\ }c }$\rightarrow$\textit{q} ($\Xi^{\ast}$) and
\textit{q }$\rightarrow$\textit{c }($(\Xi^{\ast})^{-1}$). This is an
expression of a categorical adjunction at the level of states.

\section{Selection criterion for non-equilibrium states}

The next step is to examine how to characterize a non-equilibrium state
$\omega\notin\mathcal{K}$ so as for the possibility of thermal interpretations
to be extended to the outside of $\mathcal{K}$.

\textbf{\textit{Selection criterion and thermal interpretation of
non-equilibrium local states in terms of hierarchized zeroth law of local
thermodynamics} }\cite{Oji02}: To meet simultaneously the two requirements of
characterizing an unkown state $\omega$ as a non-equilibrium local state and
of establishing its thermal interpretations, we now compare $\omega$ with
thermal reference states $\in\mathcal{K}=\Xi^{\ast}(M_{1}(V_{+}))$ by means of
local thermal observables $\mathcal{T}_{x}$ at $x$ whose thermal meanings are
provided by the corresponding thermal functions $\Xi(\mathcal{T}_{x})$ as seen
above. In view of the above conclusion [\textit{q}$\rightarrow$\textit{c}
channel $(\Xi^{\ast})^{-1}$ = thermal interpretation of quantum states] and
also of the hierarchical structure in $T_{x}$, we relax the requirement for
$\omega$ to agree with $\exists\omega_{B}:=\Xi^{\ast}(\rho_{x})\in\mathcal{K}$
up to some suitable \textit{sub}space $\mathcal{S}_{x}$ of all local thermal
observables $\mathcal{T}_{x}$. Then, we characterize $\omega$ as a
non-equilibrium local state by

\begin{itemize}
\item[iii)] a \textit{selection criterion }for $\omega$ to be $\mathcal{S}%
_{x}$\textit{-thermal} at $x$, requiring the existence of $\rho_{x}\in
M_{1}(V_{+})$ s.t.
\begin{equation}
\omega(\hat{\Phi}(x))=\Xi^{\ast}(\rho_{x})(\hat{\Phi}(x))\ \text{ for }%
\forall\hat{\Phi}(x)\in\mathcal{S}_{x},\label{S_therm}%
\end{equation}
or, $\omega\equiv\Xi^{\ast}(\rho_{x})$ (mod $\mathcal{S}_{x}$) for short. This
means that, as far as the gross thermal properties described by a subspace
$\mathcal{S}_{x}$ of $\mathcal{T}_{x}$ are concerned, $\omega$ can be
identified with a thermal reference state $\Xi^{\ast}(\rho_{x})\in\mathcal{K}%
$. The deviations of $\omega$ from $\Xi^{\ast}(\rho_{x})$ in the finer
resolutions will characterize the extent to which $\omega$ is
\textit{non-equilibrium} even locally, and, for this purpose, the hierarchical
structure inherent to $\mathcal{T}_{x}$ plays important roles. In terms of
thermal functions $\Phi:=\Xi(\hat{\Phi}(x))\in\Xi(\mathcal{S}_{x})$, the above
(\ref{S_therm}) is rewritten as
\begin{equation}
\omega(\Phi)(x):=\omega(\hat{\Phi}(x))=\rho_{x}(\Phi),\quad\Phi\in
\Xi(\mathcal{S}_{x}),\label{ThermalFn2}%
\end{equation}
which allows $\omega\equiv\Xi^{\ast}(\rho_{x})$ (mod $\mathcal{S}_{x}$) to be
solved in $\rho_{x}$ as$\ ``(\Xi^{\ast})^{-1}"(\omega)=\rho_{x}$ (mod
$\Xi(\mathcal{S}_{x})$), providing

\item[iv)] local thermal interpretations of a non-equilibrium local state
$\omega$ by means of the conditional \textit{q}$\rightarrow$\textit{c channels
}$(\Xi^{\ast})^{-1}$, conditionally meaningful on the selected states
\cite{Oji02}. We denote this similarly to (\ref{adjunction1}) by%
\begin{equation}
\mathcal{K}(\omega,\Xi^{\ast}(\rho))/\mathcal{S}_{x}\overset{q\rightleftarrows
c}{\simeq}Th((\Xi^{\ast})^{-1}(\omega),\rho)/\Xi(\mathcal{S}_{x}%
).\label{adjunction2}%
\end{equation}
\end{itemize}

The reason for mentioning here the expression, \textit{hierarchized zeroth law
of local thermodynamics}, is that the above equalities presuppose some
measuring processes of local thermal observables which involve and require the
\textit{contacts of two bodies}, measured object(s) and measuring device(s),
in a local thermal equilibrium, conditional with respect to $\mathcal{S}_{x}$.
The transitivity of this contact relation just corresponds to the localized
and hierarchized version of the standard zeroth law of thermodynamics. In this
way, we see that a selection criterion suitably set up can give a
characterization of certain class of states and, at the same time, provide
associated relevant physical interpretations of the selected states.

\section{Reformulation of DHR-DR superselection theory}

According to the above viewpoint, we can now reformulate the essence of the
DHR- and DR-superselection theory whose essence can be summarized as follows.
Here the basic ingredients of the theory with localizable charges \cite{DHR,
Haag} are a \textit{net} $\mathcal{K}\ni\mathcal{O}\longmapsto\mathfrak{A}%
(\mathcal{O})$ of local W*-subalgebras $\mathfrak{A}(\mathcal{O}) $ of
\textit{local observables}, defined on the set, $\mathcal{K}:=\{(a+V_{+}%
)\cap(b-V_{+});a,b\in\mathbb{R}^{4}\}$, of all double cones in the Minkowski
spacetime $\mathbb{R}^{4}$; it is assumed to satisfy the isotony
$\mathcal{O}_{1}\subset\mathcal{O}_{2}\Longrightarrow\mathfrak{A}%
(\mathcal{O}_{1})\subset\mathfrak{A}(\mathcal{O}_{2})$, allowing the global
(or, quasi-local) algebra of observables $\mathfrak{A}:=C^{\ast}$-
$\underset{\mathcal{K\ni O}\rightarrow\mathbb{R}^{4}}{\underset
{\longrightarrow}{\lim}}\mathfrak{A}(\mathcal{O})$ to be defined as the
C*-inductive limit, to transform covariantly under the action of the
Poincar\'{e} group $\mathcal{P}_{+}^{\uparrow}:=\mathbb{R}^{4}\rtimes
L_{+}^{\uparrow}\ni(a,\Lambda)\longmapsto\alpha_{(a,\Lambda)}\in
Aut(\mathfrak{A)}$, $\alpha_{(a,\Lambda)}(\mathfrak{A}(\mathcal{O}%
))=\mathfrak{A}(\Lambda(\mathcal{O)}+a)$, and to satisfy the local
commutativity, $[\mathfrak{A}(\mathcal{O}_{1}),\mathfrak{A}(\mathcal{O}%
_{2})]=0$ for $\mathcal{O}_{1},\mathcal{O}_{2}\in\mathcal{K}$ spacelike
separated: $\forall x\in\mathcal{O}_{1},\forall y\in\mathcal{O}_{2}$,
$(x-y)^{2}<0$.

\begin{itemize}
\item A physically relevant state $\omega\in E_{\mathfrak{A}}$ around a pure
vacuum state $\omega_{0}\in E_{\mathfrak{A}}$ is characterized by the
Doplicher-Haag-Roberts (DHR)\textit{\ }\textbf{selection criterion} requiring
for $\omega$ the existence of $\exists\mathcal{O\in K}$ s.t. $\pi_{\omega
}\upharpoonright_{\mathfrak{A}(\mathcal{O}^{\prime})}\cong\pi_{\omega_{0}%
}\upharpoonright_{\mathfrak{A}(\mathcal{O}^{\prime})}$, or, $\omega
\upharpoonright_{\mathfrak{A}(\mathcal{O}^{\prime})}=\omega_{0}\upharpoonright
_{\mathfrak{A}(\mathcal{O}^{\prime})}$, where $\mathcal{O}^{\prime}%
:=\{x\in\mathbb{R}^{4}$; $(x-y)^{2}<0$ for $\forall y\in\mathcal{O}\}$ is the
causal complement of $\mathcal{O}$ and $\mathfrak{A}(\mathcal{O}^{\prime
}):=C^{\ast}$-$\underset{\mathcal{K\ni O}_{1}\subset\mathcal{O}^{\prime}%
}{\underset{\longrightarrow}{\lim}}\mathfrak{A}(\mathcal{O}_{1})$. We denote
this criterion as
\begin{equation}
\omega\equiv\omega_{0}\text{ \ (mod. }\mathfrak{A}(\mathcal{O}^{\prime
})\text{)} .\label{DHRcrit}%
\end{equation}

\item In the GNS-vacuum representation $(\pi_{0},\mathfrak{H}_{0})$
corresponding to $\omega_{0}$, the validity of Haag duality,
\begin{equation}
\pi_{0}(\mathfrak{A}(\mathcal{O}^{\prime}))^{\prime}=\pi_{0}(\mathfrak{A}%
(\mathcal{O}))^{\prime\prime},
\end{equation}
is assumed. On the basis of the standard postulates \cite{DHR}, the selection
criterion (\ref{DHRcrit}) can be shown to be equivalent to the existence of a
\textit{local endomorphism} $\rho\in End(\mathfrak{A})$ such that
$\omega=\omega_{0}\circ\rho$, localized in some $\mathcal{O}\in\mathcal{K}$ in
the sense of%
\begin{equation}
\rho(A)=A\text{ \ \ \ for }\forall A\in\mathfrak{A(}\mathcal{O}^{\prime}).
\end{equation}

\item \textbf{DR-category }\cite{DR89}: Then, a \textbf{C*-tensor category}
$\mathcal{T}$ which we call here a DR-category is defined as a full
subcategory of $End(\mathfrak{A})$ consisting of such $\rho$'s as
\textit{objects} together with the intertwiners $T\in\mathfrak{A}$ from a
$\rho$ to another $\sigma$ s.t. $T\rho(A)=\sigma(A)T$ as \textit{morphisms}.
\newline Up to the technical details, the essence of Doplicher-Roberts
superselection theory can be summarized in the following basic results due to
the structure of $\mathcal{T}$ as a C*-tensor category\ equipped with the
permutation symmetry, the operations of taking direct sums, subobjects and
determinants, and the dominance of objects with determinant $1$:

\begin{itemize}
\item Unique existence of an \textit{internal symmetry group} $G$ such that
\begin{equation}
\mathcal{T}\simeq Rep_{G}\underset{\text{ \ \ \ \ \ \ \ \ \ \ \ \ \ \ \ \ }%
}{\overset{\text{Tannaka-Krein duality}}{\longleftrightarrow}}G=End_{\otimes
}(V),
\end{equation}
where $End_{\otimes}(V)$ is defined as the group of natural unitary
transformations $g=(g_{\rho})_{\rho\in\mathcal{T}}:V\overset{\cdot
}{\rightarrow}V$ from the C*-tensor functor $V:\mathcal{T}\hookrightarrow Hilb
$ to itself \cite{DR89, MacL}:
\begin{equation}%
\begin{array}
[c]{ccccc}%
\rho_{1} &  & V_{\rho_{1}} & \overset{g_{\rho_{1}}}{\rightarrow} & V_{\rho
_{1}}\\
T\downarrow &  & T\downarrow & \circlearrowleft & \downarrow T\\
\rho_{2} &  & V_{\rho_{2}} & \overset{g_{\rho_{2}}}{\rightarrow} & V_{\rho
_{2}}%
\end{array}
.
\end{equation}
Here, $V$ embeds $\mathcal{T}$ into the category $Hilb$ of Hilbert spaces and
its image turns out to be just the category $Rep_{G}$ of unitary
representations $(\gamma,V_{\gamma})$ of a compact Lie group $G\subset SU(d)$,
where the dimensionality$\ d$ is intrinsically defined by $\mathcal{T}$ by the
generating element $\rho\in\mathcal{T}$ \cite{DR89}. In this formulation, the
essence of the Tannaka-Krein duality \cite{TK} is found in the one-to-one
correspondence,
\begin{equation}
\mathcal{T}\ni\rho=\rho_{\gamma}\longleftrightarrow\gamma=\gamma_{\rho}%
\in\hat{G},\label{Isom}%
\end{equation}
for $\rho\in\mathcal{T}$ satisfying $\rho(\mathfrak{A})\cap\mathfrak{A}%
=\mathbb{C}\mathbf{1}$ (corresponding to the irreducibility of $\gamma_{\rho}%
$), and the identification $g_{\rho}=\gamma_{\rho}(g)$ for $g\in G$.

\item Unique existence of a \textit{field algebra}
\begin{align}
\mathfrak{F}:=  & \mathfrak{A}\underset{\mathcal{O}_{d}^{G}}{\otimes
}\mathcal{O}_{d}\text{ \ \ \ }\curvearrowleft G=Aut_{\mathfrak{A}%
}(\mathfrak{F})=Gal(\mathfrak{F}/\mathfrak{A})\\
\text{ \ \ \ \ \ \ }  & :=\{\tau\in Aut(\mathfrak{F});\tau(A)=A,\text{
}\forall A\in\mathfrak{A}\}\text{ (: Galois group),}\nonumber
\end{align}
with $\mathfrak{A}=\mathfrak{F}^{G}$ (fixed-point algebra), where
$\mathcal{O}_{d}$ is the Cuntz algebra \cite{Cuntz} defined as the unique
simple C*-algebra consisting of $d$ isometries $\psi_{i}$, $i=1,2,\cdots,d$,
\begin{equation}
\psi_{i}^{\ast}\psi_{j}=\delta_{ij}\mathbf{1,}\text{ \ \ \ }\sum_{i=1}^{d}%
\psi_{i}\psi_{i}^{\ast}=\mathbf{1,}%
\end{equation}
whose fixed-point subalgebra $\mathcal{O}_{d}^{G}$ is embedded into
$\mathfrak{A}$, $\mu:\mathcal{O}_{d}^{G}\hookrightarrow\mathfrak{A}$,
satisfying the relation $\mu\circ\sigma=\rho\circ\mu$ with respect to the
canonical endomorphism $\sigma$ of $\mathcal{O}_{d}$: $\sigma(C):=\sum
_{i=1}^{d}\psi_{i}C\psi_{i}^{\ast}$ for $C\in$ $\mathcal{O}_{d}$.
\end{itemize}

\item The \textit{superselection structure} in the irreducible vacuum
representation $(\pi,\mathfrak{H})$ of the constructed field algebra
$\mathfrak{F}$ is understood as follows: first, the group $G$ of symmetry
arising in this way is \textbf{unbroken} with a unitary implementer
$U:G\rightarrow\mathcal{U}(\mathfrak{H}),$ $\pi(\tau_{g}(F))=U(g)\pi
(F)U(g)^{\ast}$ and is \textit{global }(i.e., gauge symmetry of the 1st kind)
[due to the transportability in spacetime imposed on each $\rho\in\mathcal{T}%
$]. $\mathfrak{H}$ contains the starting Hilbert space $\mathfrak{H}_{0}$ of
the vacuum representation $\pi_{0}$ of $\mathfrak{A}$ as a cyclic $G$-fixed
point subspace, $\mathfrak{H}_{0}=\mathfrak{H}^{G}=\{\xi\in\mathfrak{H}$;
$U(g)\xi=\xi$ for $\forall g\in G\}$, $\overline{\pi(\mathfrak{F}%
)\mathfrak{H}_{0}}=\mathfrak{H}$. Then, $\mathfrak{H}$ is decomposed into a
direct sum in the following form \cite{DHR},
\begin{align}
\mathfrak{H}  & =\underset{\gamma\in\hat{G}}{\oplus}(\mathfrak{H}_{\gamma
}\otimes V_{\gamma}),\\
\pi(\mathfrak{A})  & =\underset{\gamma\in\hat{G}}{\oplus}(\pi_{\gamma
}(\mathfrak{A})\otimes\mathbf{1}_{V_{\gamma}}),\text{ \ \ \ }U(G)=\underset
{\gamma\in\hat{G}}{\oplus}(\mathbf{1}_{\mathfrak{H}_{\gamma}}\otimes
\gamma(G)),\label{sector}%
\end{align}
where \textbf{superselection sectors} defined as mutually disjoint irreducible
representations $(\pi_{\gamma},\mathfrak{H}_{\gamma})$ of $\mathfrak{A}$ are
in one-to-one correspondence, $\pi_{\gamma}=\pi_{0}\circ\rho_{\gamma
}\longleftrightarrow\rho_{\gamma}\in\mathcal{T}$ $\longleftrightarrow
(\gamma,V_{\gamma})$, with mutually disjoint irreducible unitary
representations $(\gamma,V_{\gamma})$ of $G$, the totality of which just
constitutes the group dual $\hat{G}$.
\end{itemize}

Now, introducing the map $W$ defined by
\begin{equation}
W:End(\mathfrak{A})\ni\rho\longmapsto\omega_{0}\circ\rho\in E_{\mathfrak{A}},
\end{equation}
we reformulate the essence of the above DHR-DR theory in a form parallel to
the previous sections. First, the role of the DHR criterion is easily seen to
make the map $W:End(\mathfrak{A})\rightarrow E_{\mathfrak{A}}$
\textit{invertible }on the subset $E_{DHR}$ selected by it:
\begin{align}
E_{DHR}  & :=\{\omega\in E_{\mathfrak{A}};\exists\mathcal{O\in K}\text{ s.t.
}\omega\equiv\omega_{0}\ (\text{mod.}\mathfrak{A}(\mathcal{O}^{\prime
}))\}\nonumber\\
& \ni\omega=\omega_{0}\circ\rho\longmapsto\rho\in\mathcal{T\subset
}End(\mathfrak{A}),\label{DHRcrit1}%
\end{align}
which yields an alternative description of states $\omega\in E_{DHR}$ in terms
of C*-tensor category $\mathcal{T}$ of local endomorphisms. What is important
about (\ref{sector}) is that it means the existence of a non-trivial centre of
$\pi(\mathfrak{A})^{\prime\prime}$,
\begin{equation}
\mathfrak{Z}(\pi(\mathfrak{A})^{\prime\prime})=\underset{\gamma\in
\hat{G}}{\oplus}\mathbb{C}(\mathbf{1}_{\mathfrak{H}_{\gamma}}\otimes
\mathbf{1}_{V_{\gamma}})=C(\hat{G})=\mathfrak{Z}(U(G)^{\prime\prime
}).\label{centre}%
\end{equation}
In view of the one-to-one correspondence (\ref{Isom}) due to the isomorphism,
$\mathcal{T}\simeq Rep_{G}$, we define a unital positive linear map
$k:\mathfrak{A}\ni A\longmapsto k(A)\in\mathfrak{Z}(\pi(\mathfrak{A}%
)^{\prime\prime})=C(\hat{G})$\ by
\begin{equation}
\lbrack k(A)](\gamma):=\omega_{0}(\rho_{\gamma}(A))=W(\rho_{\gamma})(A),
\end{equation}
whose positivity and the normalization are evident: $[k(A^{\ast}A)](\gamma
)=$\newline $=\omega_{0}(\rho_{\gamma})(A^{\ast}A)\geq0$, $[k(\mathbf{1}%
)](\gamma)=\omega_{0}(\rho_{\gamma})(\mathbf{1})=1$. Thus, similarly to the
\textit{c}$\rightarrow$\textit{q} channel $\Xi^{\ast}$ in Sec.2, the dual of a
completely positive (CP) map $k$ defines a \textit{c}$\rightarrow$\textit{q}
channel $k^{\ast}:M_{1}(\hat{G})\rightarrow E_{\mathfrak{A}}$, given for each
probability weight $\nu=(\nu_{\gamma})_{\gamma\in\hat{G}}\in M_{1}(\hat{G})$,
$\nu_{\gamma}\geq0$, $\sum_{\gamma\in\hat{G}}\nu_{\gamma}=1$, by
\begin{align}
& k^{\ast}(\nu)(A)=\nu(k(A))=\sum_{\gamma\in\hat{G}}\nu_{\gamma}%
[k(A)](\gamma)\nonumber\\
& =\sum_{\gamma\in\hat{G}}\nu_{\gamma}\omega_{0}(\rho_{\gamma}(A))=(\sum
_{\gamma\in\hat{G}}\nu_{\gamma}\omega_{\gamma})(A)\nonumber\\
& \Longrightarrow k^{\ast}(\nu)=\sum_{\gamma\in\hat{G}}\nu_{\gamma}%
\omega_{\gamma}=\sum_{\gamma\in\hat{G}}\nu_{\gamma}W(\rho_{\gamma}),
\end{align}
where $\omega_{\gamma}:=\omega_{0}\circ\rho_{\gamma}$. Therefore, the map
$k^{\ast}$ extends $W$ to ``convex combinations'' of $\rho_{\gamma}$'s, and is
a ``\textit{charging} map'' to create from the vacuum $\omega_{0} $ a state
$k^{\ast}(\nu)=\sum_{\gamma\in\hat{G}}\nu_{\gamma}(\omega\circ\rho_{\gamma})$
whose charge contents are described by the charge distribution $\nu
=(\nu_{\gamma})_{\gamma\in\hat{G}}\in M_{1}(\hat{G})$ over the group dual
$\hat{G}$. While this picture clearly shows the parallelism with the previous
discussion of the thermal interpretation based upon the \textit{c}%
$\rightarrow$\textit{q }channel\textit{\ }$\Xi:\mathcal{A}\rightarrow
C(V_{+})$, a different but equivalent formulation will be more familiar for
describing the same situation in use of a reducible representation
$\gamma_{\nu}:=\oplus_{\gamma\in\hat{G},\nu_{\gamma}\neq0}\gamma\in RepG$ as
follows. For this purpose, we use the invariance of the vacuum state under
$U(G)$ which implies the following relations in terms of the conditional
expectation $m:\mathfrak{F}\ni F\longmapsto m(F):=\int_{G}dg\tau_{g}%
(F)\in\mathfrak{A}$ coming from the average over $G $ by the Haar measure $dg$
:
\begin{equation}
\omega_{\gamma}(m(F))=(\omega_{0}\circ\rho_{\gamma})(m(F)))=\langle\Omega
_{0}\ |\ \sum_{i}\psi_{i}^{\gamma}m(F)\psi_{i}^{\gamma\ast}\Omega_{0}\rangle,
\end{equation}
where the last expression is understood in the representation space
$\mathfrak{H}$ of $\mathfrak{F}$ and $\psi_{i}^{\gamma}\in\mathfrak{F}$ are
such that $\psi_{i}^{\gamma}\pi(A)=\pi\circ\rho_{\gamma}(A)\psi_{i}^{\gamma}$
for $\forall A\in\mathfrak{A}$. Then, owing to the disjointness among
different sectors, the state $k^{\ast}(\nu)$ can be rewritten as an induced
state $k^{\ast}(\nu)\circ m$ of $\mathfrak{F}$ by
\begin{align}
k^{\ast}(\nu)(m(F))  & =\sum_{\gamma\in\hat{G}}\nu_{\gamma}\omega_{\gamma
}(m(F))=\sum_{\gamma\in\hat{G}}\sum_{i}\langle\sqrt{\nu_{\gamma}}\psi
_{i}^{\gamma\ast}\Omega_{0}\ |\ m(F)\sqrt{\nu_{\gamma}}\psi_{i}^{\gamma\ast
}\Omega_{0}\rangle\nonumber\\
& =\langle\Psi\ |\ m(F)\Psi\rangle=\langle\Psi\ |\ F\Psi\rangle,
\end{align}
with a vector $\Psi:=\sum_{\gamma\in\hat{G}}\sum_{i}\sqrt{\nu_{\gamma}}%
\psi_{i}^{\gamma\ast}\Omega_{0}\in\mathfrak{H}$ belonging to the above
mentioned reducible representation $\gamma_{\nu}:=\oplus_{\gamma\in
\hat{G},\nu_{\gamma}\neq0}\gamma$ of $G$. Thus, on the basis of the basic
results of DHR-DR theory, we have the inverse map $(k^{\ast})^{-1}$ on
$E_{DHR}$ as a \textit{q}$\rightarrow$\textit{c} channel, which provides the
interpretation of a state $\omega\in$ $E_{DHR}$ selected out by the DHR
criterion, with respect to their internal-symmetry aspects, specifying its $G
$\textit{\textbf{-charge contents}} understood as the $G$-representation
contents. Since the spacetime behaviours of quantum fields are expressed by
the observable net $\mathcal{O}\longmapsto\mathfrak{A}(\mathcal{O})$ and since
the internal symmetry aspects are described in the above machinery also
encoded in $\mathfrak{A}$, the role of the field algebra $\mathfrak{F}$ and
the internal symmetry group $G$ seems to be quite subsidiary, simply providing
comprehensible vocabulary based on the covariant objects under the symmetry transformaions.

According to the naive physical picture, contents of the sectors parametrized
by $\gamma\in\hat{G}$, $\gamma\neq\iota$(: the trivial representation
corresponding to the vacuum sector) are \textit{excited states }above the
vacuum. What is lacking in the stardard theory of superselection sectors is
this kind of problems concerning the \textit{\textbf{energy contents of
sectors}}, namely, the mutual relations between the energy-momentum spectrum
and the sectors as internal-symmetry spectrum: since only the sector of the
trivial representation $\iota\in\hat{G}$ contains the vacuum state with the
\textit{minimum energy} $0$ and since all other sectors consist of the excited
states, the above picture suggests the following situations to be valid:
\begin{equation}
\min\{Spec(\hat{P}_{0}\upharpoonright_{\mathfrak{H}_{0}})\}=0,\text{
\ \ \ }\inf\{Spec(\hat{P}_{0}\upharpoonright_{\mathfrak{H}_{0}^{\perp}})\}>0.
\end{equation}
In the treatment of thermal functions in Sec.2, it is easily seen that, while
the entropy density $s(\beta)$ is not contained in the image set
$\Xi(\mathcal{T}_{x})$ due to the absence of such a quantum observable
$\hat{s}(x)\in\mathcal{T}_{x}$ that $\omega_{\beta}(\hat{s}(x))=s(\beta)$, it
can be approximated by the thermal functions belonging to $\Xi(\mathcal{T}%
_{x})$. In order to facilitate the discussions of problems of this sort, it is
important to have those observables freely at hand which detect the
$G$-charge\textit{\textbf{\ }}contents in $\mathfrak{Z}(\pi(\mathfrak{A}%
)^{\prime\prime})=C(\hat{G})$, and, for this purpose, we need also here to
consider the problem as to how such observables can be supplied from the
\textit{local} observables belonging to $\mathfrak{A}(\mathcal{O})$, i.e., the
approximation of order parameters by order fields or central sequences. For
this purpose, the analyses of point-like fields and the rigorous method of
their operator-product expansions developed by Bostelmann in \cite{Bo} would
be quite useful in these contexts. All the above sort of considerations (with
the modifications of the DHR selection criterion necessitated by the possible
presence of the long range forces, such as of Buchholz-Fredenhagen type) will
be crucially relevant to the approach to the colour confinement problem, and,
especially the latter one seems to be quite a non-trivial issue there.

\section{Selection criteria as categorical adjunctions}

Here we emphasize the important roles played by the \textit{categorical
adjunctions} underlying our discussions so far, in achieving the systematic
organizations of various different theories in physics, controlled by the
selection criteria to characterize specific physical domains so that the
physical interpretations are automatically afforded. While the two formal
formulae (\ref{adjunction1}) and (\ref{adjunction2}) mentioned at the end of
Sec.2 and Sec.3 were just for the sake of suggesting such relevance of the
adjunctions, our selection criteria, expressed in terms of states, may not
directly look like a familiar form of categorical adjunctions \cite{MacL}:%
\begin{equation}
A(F(x),a)\simeq X(x,G(a)),\label{adjunction3}%
\end{equation}
involving two functors, a forgetful functor $G:A\rightarrow X$, and its left
adjoint $F:X\rightarrow A$. This is mainly because the category consisting of
states of (C*-)algebras is a rather rigid one, allowing only few meaningful
morphisms among different objects, requiring a strict equality, or, at most,
some closeness w.r.t. a norm or seminorms. For this reason, the essence of
adjunctions in our context mostly is found in such a quantitative form as
\begin{equation}%
\begin{array}
[c]{ccc}%
X\text{ (= q): to be classified} & q\leftarrow c & A\text{ (= c): to
classify}\\
x\equiv G(a)\text{ (mod.}X\text{)} & \underset{F}{\overset{G}{\leftrightarrows
}} & F(x)\equiv a\text{ (mod.}A\text{)}\\
\text{as selection criterion} & q\rightarrow c & \text{interpretation}%
\end{array}
,
\end{equation}
with $A$ a classical classifying space as the spectrum of centre, $X$ a
quantum domain (of states) to be classified, $G$ the \textit{c}$\rightarrow
$\textit{q channel} and with $F$ the \textit{q}$\rightarrow$\textit{c channel}
to provide the interpretation of $X$ in terms of the vocabulary in $A$.

As we have seen above, once the contents of imposed selection criteria are
paraphrased into different languages, such as thermal functions in Sec.2 and
local endomorphisms in Sec.4, then the standard machinery stored in the
category theory starts to work, in such a form as intertwiners among local
endomorphisms or among group representations, which provides us with powerful
tools to analyze the given structures, as seen in Sec.4 and II; Sec.3.

For decoding the deep messages encoded in a selection criterion, what plays
the decisive roles at the first stage is the identification of the
\textit{centre} of a representation containing universally all the selected
quantum states; its spectrum provides us with the information on the structure
of the associated superselection sectors, which serves as the vocabulary to be
used when the interpretations of a given quantum state are presented. The
necessary bridge between the selected generic quantum states and the classical
familar objects living on the above centre is provided, in one direction, by
the \textit{c}$\rightarrow$\textit{q channel} which embeds all the known
classical states (=probability measures) into the form of quantum states
constituting the totality of the selected states by the starting selection
criterion. The achieved identification between what is selected and what is
embedded from the known world is nothing but the most important consequence of
the categorical adjunction formulated in the form of selection criterion. This
automatically enables us to take the inverse of the \textit{c}$\rightarrow
$\textit{q channel}\ which brings in another most important ingredient, the
\textit{q}$\rightarrow$\textit{c channel} to decode the physical contents of
selected states from the viewpoint of those aspects selected out by the
starting criterion. Mathematically speaking, the spectrum of the above centre
is nothing but the \textit{classifying space }universally appearing in the
geometrical contexts, for instance, in Sec.4 of DR superselection theory of
unbroken symmetry described by a compact Lie group $G$, its dual $\hat{G}$ (of
all the irreducible unitary representations) is such a case,
$\hat{G}=B_{\mathcal{T}}$ for $\mathcal{T}$ the DR category of local
endomorphisms of the observable net, where our \textit{q}$\rightarrow
$\textit{c channel }$(k^{\ast})^{-1}$ plays the role of the
\textit{classifying map} by embedding the $G$-representation contents of a
given quantum state into the subset of $\hat{G}$ consisting of its irreducible
components:
\begin{equation}%
\begin{array}
[c]{ccc}%
\underset{\gamma\in M}{\amalg}V_{\gamma}=E &  & (\pi_{u},\underset{\gamma
\in\hat{G}}{\oplus}V_{\gamma})\text{: all the sectors}\\
\text{ \ \ \ \ \ \ \ \ \ \ \ \ \ \ \ }\downarrow_{RepG} &  & \downarrow
_{RepG}\text{ \ \ \ \ \ \ \ \ \ \ \ }\\
(RepG\supset)M & \overset{(k^{\ast})^{-1}}{\hookrightarrow} &
\hat{G}=B_{\mathcal{T}}\text{ \ \ \ \ \ \ \ \ \ \ \ \ \ \ \ }%
\end{array}
.
\end{equation}
From this viewpoint, the present scheme can easily be related with many
current topics concerning the geometric and classification aspects of
commutative as well as non-commutative geometry based upon the (homotopical)
notions of classifying spaces, K-theory and so on.

\end{document}